# On the Boundaries of Trust and Security in Computing and Communications Systems

## Al-Sakib Khan Pathan

Department of Computer Science
International Islamic University Malaysia
Gombak 53100, Kuala Lumpur, Malaysia
Phone: +603-61964000 Ext. 5653
E-Mail: sakib@iium.edu.my, sakib.pathan@gmail.com

**Abstract:** This article analyzes trust and security in computing and communications systems. While in human-life, trust usually has some kind of commonly understood meaning, in the realm of computing and communications systems, it could be interpreted differently in different environments and settings. On the other hand, security is about making sure that the participating entities are legitimate in a communication event or incident so that the core requirements of privacy, integrity, and authenticity are maintained. This notion is also true for our human life, even for example entering a house needs legitimacy of a person. Some boundary lines preserve the security; otherwise an unwanted access is called a '*security breach*'. The intent of this article is to compare and discuss these two terms with our societal behavior and understanding amongst entities. To illustrate these issues especially in computing and communications world, some of the innovating and recent technologies are discussed which demand trust and security within their core operational structures. Alongside presenting generally established ideas, some critical points are mentioned that may be sometimes debatable within the research community.

## 1 The Notions of Trust and Security

It is a debatable issue whether trust should be considered fully within the perimeter of security in computing and communications systems. In usual human life, we see these two terms go side-by-side to define the relationships we might have with other fellow human beings. Trust among people sets the level of security felt by each person involved in various relationships. If person A does not trust person B, person A may not feel secure with the company of person B. Similarly, these terminologies could also retain same meanings for our modern computing and communications equipments. However, a notable difference is that while trust and security are very interrelated in human-life scenario; in technical fields, these are considered as two clearly different issues with clear boundary lines. In computing and communications fields, trust is a kind of vague term that sets the outline of a task or communication event, based on which the operation can be performed. On the other hand, security is a broad concept that ensures the communications go forward in a desired way maintaining the core requirements of security intact, i.e., privacy, authenticity, authority, integrity, and non-repudiation. The relation between trust and security could be seen in the way that security includes trust concept partially and trust stays as a wrapper before any secure or insecure communication happens within a network of devices.

As we are moving towards a big switch in Information Technology (IT) world with the developments of innovative communications and computing technologies, managing trust among the participating entities has become one of the major concerns today. The entities may mean different types of technical devices as well as the users or people who are associated with these devices (or, who are taking advantage of these devices for their daily interactions among themselves and other species). To initiate any communication, it is considered that the initiator already gives a trust value to the entity that is communicated. Based on the acceptable reply, the communication door may be opened or may be terminated; which means, either that initially put trust is established or broken afterwards. If the trust is maintained by both entities and they enter into any secure communication using some kind of mechanism, we say that both the entities have used trust to achieve secure communication as well as security of the communication has ensured a better trust level among the participants. In real life, it may take a long time to establish trust among entities based on some kind of knowledge-base about the participating entities. A suspicious action by any entity may threaten long-established trust; again a proven wrong action of an entity may completely tear down long-established trust.

## 2  Reputation-based Trust and Security

As establishing trust needs longtime-observed behavioral analysis of an entity (in most of our perceived domains), a commonly used technique is to give some kind of reputation value (Shmatikov and Talcott, 2005) to each participating entity in a system. This value increases with the reliable contributions of the entity (i.e., a device or equipment or a user or a client) in the system, while decreases with wrong or suspicious behavior. Reputation could be analyzed using various underlying methods. Sometimes the peer-information is used, sometimes some kind of regional information is kept, sometimes for smaller scale systems a global knowledge is used for maintaining the reputation values, and so on. Various works have mentioned using game-theory (Seredynski and Bouvry, 2009), probability model (Dong et al., 2009), fuzzy logic (Li et al., 2010) and other techniques (e.g., distributed reputation management (Bamasak and Zhang, 2005), using reputation matrix (Wei et al., 2012)) for reputation analysis.

Like the reputation-based trust, reputation-based security uses similar idea of keeping reputation of the computing and communication devices at an acceptable level but the difference is that it is channeled towards maintaining protection that is sufficient to meet the aspects of security in the system's operational level. Often trust remains as a wrapper and security is more meaningful for the actual implementation of taking some steps for communications using the reputation that is gained.

## 3  Trust and Security in Prominent Future Technologies

### 3.1  Pervasive and Ubiquitous Computing

Ubiquitous Computing or Pervasive Computing is used interchangeably to label the use of computing technologies and devices in daily human life. The basic idea is to make technology easily accessible to human beings instead of coercing them to adapt the technological advancements. This field basically stems from Human-Computer Interaction (HCI) field, then got support from other fields and complex blend of various other areas contributed to its development. If technical devices blend with usual human life, trusting the objects within the environment comes forward as an issue of paramount importance. Such trust among the electronic/technical devices can be achieved through reputation based systems as discussed earlier.

On the other hand, security in case of pervasive computing is often difficult to define as the reality is; it is possible to give a sense of security in pervasive computing environment (Pagter and Petersen, 2007) but as the devices could be of different platforms and features, it is really a complicated task to ensure proper security in pervasive communications.

*3.2 Internet of Things and Future Internet*

The Internet of Things (IoT) is the terminology used for uniquely identifiable objects (or, things) and their virtual representations in an Internet-like structure. The idea of IoT considers same ubiquitous environment as that is for ubiquitous or pervasive computing. Basic difference between pervasive computing and IoT is that when the same setting is seen from the conceptual angle, it is termed pervasive environment but when seen from the angle of identifiable objects taking part in the system, it is called IoT, an Internet-like network of networks. As IoT and pervasive computing deal with the same environment of connected devices (i.e., different kinds of devices of diverse natures and computing powers), trust and security management are also similar to the methods discussed earlier; that means, a scheme developed for pervasive computing environment could also be applicable in most of the cases to IoT, when the device-level trust and security are dealt with.

While IoT is clearly defined by this time, Future Internet is still standing as a foggy term. Future Internet basically refers to a wide variety of research topics related to the idea of Internet of connecting numerous networking devices around the globe. If simply a relatively faster Internet with new devices and techniques is brought forward at the end, it could end up just as an extension of the current Internet that we have. However, the basic vision behind Future Internet is that it is not the Internet as we have seen so far; it may have a new way of working, it may have a new method of connecting devices, and there might be even complete clean-slate approach of developing it. As the full operational definition is not yet finalized, trust and security in Future Internet are also in the preliminary survey stage and cannot be outlined clearly.

*3.3 Wireless Ad hoc and Sensor Networks*

Wireless Ad hoc network is a combination of computing nodes that can communicate with each other without the presence of a formal central entity (infrastructure-less or semi-infrastructure based) and could be established anytime, anywhere. Each node in an Ad hoc network can take the roles of both a host and a router-like device within the network. There might be different forms of Ad hoc networks like Mobile Ad hoc Network (MANET), Vehicular Ad hoc Network (VANET), Wireless Mesh Network (WMN), Wireless Sensor Network (WSN), Body Area Network (BAN), Personal Area Network (PAN), etc. Though all of these derive some common features of Ad hoc technology, WSN is a network to mention distinctively as this type of network comes with the extra feature that it might have a base station, thus a fixed central entity for processing network packets and all other sensor nodes in the network could be deployed on ad hoc basis.

The basic characteristics of Ad hoc networking require trust for initializing the communications process. In most of the cases, some kind of assumed model is needed to introduce the notion of belief and to provide a dynamic measure of reliability and trustworthiness in the ad hoc scenario (Pirzada and McDonald, 2004). In spite of having the extra advantage of some kind of central authority in WSN (where every participating entity is not very distributed like pure Ad hoc networks), trust related research works still need concrete achievements to reach a satisfactory level. Regarding security, there are hundreds of works on key management, secure routing, security services, and intrusion detection systems for any kind of Ad hoc networks (Pathan, 2010).

*3.4 Cloud Computing*

Cloud computing is a recently coined term when distributed processing is seen from the 'computing service' providing perspective. The basic concept of Cloud computing is paying based on the resource usages and the IT (Information Technology) resources that do not exist on the users' side. This strategy makes the users capable of doing tasks with their resource-constrained computing devices that they could not have done before. To support such type of communications and computing, there are many research issues to consider so that the resources could be managed efficiently and securely. The reality is that the service dynamism, elasticity, and choices offered by this highly scalable technology are very attractive for various enterprises. These opportunities, however, put lots of challenges when trust and security are considered. Cloud computing has opened up a new frontier of challenges by introducing a different type of trust scenario (Khan and Malluhi, 2010) where the technology needs primary trust of participating entities to start functioning.

While trust may still remain relevant to the Cloud concept, security puts a major difficulty when authentication is taken into consideration (for example). The idea of Cloud is to keep the service providers' identities in a cloudy state, but authentication, as a part of security requires exact identification of the objects so that the property of non-repudiation can also be ensured. Non-repudiation, in plain term means the ability to ensure that a party in a contract or a communication event cannot deny the authenticity of its own messages, packets, signature, or any kind of document that it generates. These conflicting principles have kept Cloud computing security still as a very challenging area of research. One possible solution is to maintain some kind of trusted service-providers and using them through a trusted service-providing server or service-manager. Another question in this field is that in such setting, how long an established trust should be kept alive or how to tear it down maintaining the concept of Cloud intact (that is the clients do not know from which exact locations they are getting the required services or how their requests have been processed)?. In fact, this is an area where trust and security are very much intertwined. Elements of Cloud computing also blend into the notion of Future Internet, leading to the concept of Cloud networking. Hence, advancement in the areas of trust and security in Cloud computing may also contribute to the same areas of Future Internet.

**4 Concluding Remarks**

In today's world, big companies like Amazon, E-Bay (or similar companies) use some kind of rating system that is based on trust or that lays the foundation of trust on products sold using those channels. Security is built upon the gained trust and mainly is used for allowing monetary transactions.

Trust management in reality is a complex mixture of different fields if we consider the computing and communication technologies altogether. In this article, some future technologies are discussed but there are numerous other emerging fields like near-field communications (NFC), electronic knowledge management, nano-communication networks, etc. that will also need support of trust and security. Researchers and practitioners from various fields like networking, grid and Cloud computing, distributed processing, information systems, human computer interaction, human behavior modeling could be the contributors and combination of different fields under this umbrella will become inevitable in the coming days.

Whatever the advancements we could get in concrete scale in the upcoming days, some fundamental questions could still remain like; how long an established trust be maintained or kept valid in any computing and communication system? If periodic refreshing and re-establishing trust is needed, what could be the optimal interval for different settings? Will periodic re-establishment of trust affect the established secure channels? Will security be able to act on its own when trust parameters are completely ignored? The major questions will still remain: Will there be clear boundaries between trust and security in the future innovative technologies or should trust be considered within the perimeter of security?

# References

Bamasak, O. and Zhang, N. (2005) 'A distributed reputation management scheme for mobile agent based e-commerce applications', *The 2005 IEEE International Conference on e-Technology, e-Commerce and e-Service (EEE '05)*, pp.270-275.

Dong, P., Wang, H., and Zhang, H. (2009) 'Probability-based trust management model for distributed e-commerce', *IEEE International Conference on Network Infrastructure and Digital Content (IC-NIDC 2009)*, Beijing, China, pp.419-423.

Khan, K.M. and Malluhi, Q. (2010) 'Establishing Trust in Cloud Computing', *IT Professional*, Volume 12, Issue 5, pp.20-27.

Li, J., Liu, L., and Xu, J. (2010) 'A P2P e-Commerce Reputation Model Based on Fuzzy Logic', *2010 IEEE 10th International Conference on Computer and Information Technology (CIT)*, Bradford, UK, pp.1275-1279.

Pagter, J.I. and Petersen, M.G. (2007) 'A sense of security in pervasive computing: is the light on when the refrigerator door is closed?', *Proceedings of the 11th International Conference on Financial cryptography and 1st International conference on Usable Security (FC'07/USEC'07)*, Berlin, Heidelberg, Germany, Available at http://usablesecurity.org/papers/pagter.pdf

Pathan, A.-S.K. (2010) Security of Self-Organizing Networks: MANET, WSN, WMN, VANET. (edited and contributed), ISBN: 978-1-4398-1919-7, Auerbach Publications, CRC Press, Taylor & Francis Group, USA.

Pirzada, A.A. and McDonald, C. (2004) 'Establishing trust in pure ad-hoc networks', *Proceedings of the 27th Australasian conference on Computer Science*, Volume 26.

Seredynski, M. and Bouvry, P. (2009) 'Evolutionary game theoretical analysis of reputation-based packet forwarding in civilian mobile Ad Hoc networks', *IEEE International Symposium on Parallel & Distributed Processing (IPDPS 2009)*, Rome, Italy, pp.1-8.

Shmatikov, V. and Talcott, C. (2005) 'Reputation-Based Trust Management', *Journal of Computer Security*, Volume 13, Issue 1, pp.167-190.

Wei, X., Ahmed, T., Chen, M., and Pathan, A.-S.K. (2012) 'PeerMate: A Malicious Peer Detection Algorithm for P2P Systems based on MSPCA', *International Conference on Computing, Networking and Communications (IEEE ICNC 2012)*, Maui, Hawaii, USA. (To Appear)